\documentclass[aps,reprint,prl,superscriptaddress,floats,floatfix,nobibnotes]{revtex4-2}

\usepackage{amsmath}
\usepackage{amsfonts}
\usepackage{amssymb}
\usepackage{graphicx}
\usepackage{color}
\usepackage[colorlinks=true,citecolor=blue,linkcolor=blue,urlcolor=blue]{hyperref}
\usepackage{times}
\usepackage{amstext}
\usepackage{latexsym}
\usepackage{lipsum}
\usepackage{bm}
\usepackage{mathtools}

\usepackage{orcidlink}

\usepackage{cleveref}

\providecommand{\moy}[1]{\langle #1 \rangle}

\providecommand{\ket}[1]{\lvert #1 \rangle}

\begin{document}

\title{High Efficiency Storage of Quasi-Classical and Quantum States in Coupled Resonators}

\author{Luiz O. R. Solak\,\orcidlink{0000-0002-4760-3357}}
\email{luiz7696@gmail.com}

\affiliation{Departamento de Física, Universidade Federal de São Carlos, 13565-905 São Carlos, São Paulo, Brazil}

\affiliation{CESQ/ISIS (UMR 7006), Université de Strasbourg and CNRS, 67000 Strasbourg, France}

\author{Ciro M. Diniz\,\orcidlink{0000-0002-7602-0468}}
\affiliation{Departamento de Física, Universidade Federal de São Carlos, 13565-905 São Carlos, São Paulo, Brazil}

\author{Daniel Z. Rossatto\,\orcidlink{0000-0001-9432-1603}}
\affiliation{Universidade Estadual Paulista (UNESP), Instituto de Ciências e Engenharia, 18409-010 Itapeva, São Paulo, Brazil}

\author{Antonio S. M. de Castro\,\orcidlink{0000-0002-1521-9342}}
\affiliation{Universidade Estadual de Ponta Grossa (UEPG), Departamento de Física, 84030-900, Ponta Grossa, PR, Brazil}

\author{Charles A. Downing\,\orcidlink{0000-0002-0058-9746}}
\affiliation{Department of Physics and Astronomy, University of Exeter - Exeter EX4 4QL, UK}

\author{Celso J. Villas-Boas\,\orcidlink{0000-0001-5622-786X}}
\affiliation{Departamento de Física, Universidade Federal de São Carlos, 13565-905 São Carlos, São Paulo, Brazil}

\begin{abstract}
We propose an optical model in which both quantum and quasi-classical states can be ideally stored using coupled resonators. The protocol is based on a time-dependent coupling between two cavities, carefully modulated to allow the complete transfer of an external propagating field from one cavity to another. The system maintains high storage efficiency (above $99.99\%$) even when error sources are introduced (up to $5\%$) in the coupling, such as amplitude deviation or a time delay between field propagation and coupling control. 
Furthermore, this procedure can be extended to store entangled states by considering either a pair of systems or bimodal cavities. Due to its high efficiency, this model may find application in current quantum technologies, such as quantum memories and quantum batteries, which rely on efficient quantum state storage.
\end{abstract} 

\maketitle

\textit{Introduction}---The efficient storage of quantum states is crucial for quantum communication and quantum computation \cite{monroe2002,kimbleqi}. In that context, \textit{quantum memories} plays a pivotal role in the development of quantum-enabled technologies \cite{Simon2010, Heshami2016,lvovsky2009}. These devices require on-demand storage and retrieval of qubits encoded in photon states, generally using their interaction with matter. A wide variety of protocols were proposed over the years for different platforms such as single atoms \cite{maitre1997,Giannelli_2018,kollath_2024, arslanov2021, kharlamova2024}, atomic ensembles \cite{fleischhauer2002,Moiseev2011} that can be based on atomic gases \cite{novikova2007}, solid-state systems \cite{tittel2010,Saglamyurek2011} and by means of the electromagnetically induced transparency \cite{lukin2003,eisaman2005,Chaneliere2005,chu2024}. A consequence of using atoms as a memory medium is that atomic losses inevitably will decrease the overall efficiency. In this regard, optical memories are a source of investigation as well \cite{Pittman2002,leung2006,bouillard2019,evans_2023}, being those schemes mostly based on looping the light states for arbitrary times.

Alternatively, storing quantum states is part of a more recent and yet relevant uprising technology, the so-called \textit{quantum batteries} \cite{alicki2012,Binder2015,campaioli2017,campioli2024_rev}. These devices should store energy as their classical counterparts, but with clear advantages related to the scaling of the charging power due to the quantum nature of the charged states \cite{campaioli2017}. As important as the battery itself, the drive can highly impact its overall efficiency \cite{zhang2019,Chen2020,Crescente2020,Downing2024,gemme2024}. Most recently, an exponential enhancement in both stored energy and charging power was theoretically shown using quadratic driving fields in Gaussian envelopes \cite{downing2024b}, highlighting the importance of these already well-established light pulses. 

\begin{figure}[t]
    \includegraphics[width=\linewidth]{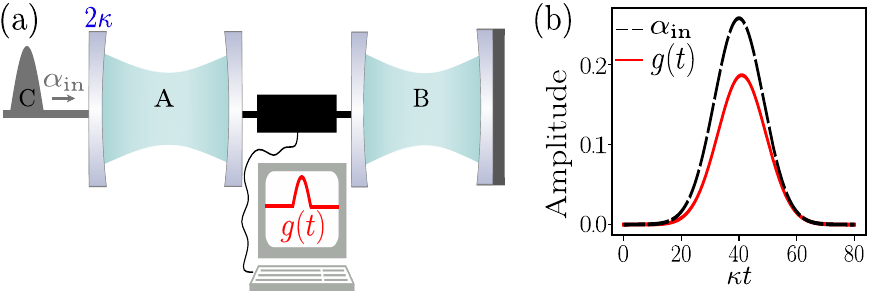}
    \caption{{Perfect state storage in coupled cavities.} (a) Pictorial representation of the physical system proposed for storage of both quasi-classical and quantum states. It is composed of two cavities $\text{A}$ and $\text{B}$ interacting via a time-dependent coupling $g(t)$ which needs to be modulated so that any square-normalized input state $\alpha_\text{in}(t)$ can be stored in the cavity mode $\text{B}$, resulting in no output state in the reservoir $\text{C}$. (b) Example of the amplitudes of both $g(t)$ and $\alpha_{\text{in}}(t)$ needed to achieve ideal storage of a single-photon state.}
    \label{model}
\end{figure}

In this work, we propose a novel, simple, and highly efficient optical model that can store quantum (single photon or entangled) and quasi-classical (coherent) states of light using coupled resonators, making this system suitable for quantum memories and potentially for quantum batteries. The protocol is theoretically straightforward, considering a system composed of coupled optical cavities $\text{A}$ and $\text{B}$ [Fig.~\ref{model}(a)]. 
In addition, we assume that the cavity mode $\text{A}$ is coupled to a bosonic reservoir $\text{C}$, on which the light states will be impinging. Lastly, we assume phase-match between the input light pulse and intracavity field, resulting in no output field, whose assumption is adapted from the condition originally proposed for complete absorption of a photon in atom-cavity systems in Ref.\cite{kuhn2012}. Then, by controlling the time-dependent coupling $g(t)$ between $\text{A}$ and $\text{B}$ [Fig.~\ref{model}(b)], the input state can be fully stored in the cavity mode $\text{B}$, in an ideal scenario, depending solely on the temporal shape of the input pulse $\alpha_\text{in}(t)$. 
Outside of the ideal case, we introduce possible sources of error to the system, such as coupling amplitude deviation and a delay between the coupling and the input state. Even in such cases, we found that the model still has high storing efficiency ($>99.99\%$) for amplitude deviation ($ \lesssim 5\%$) and time delay ($ \lesssim 5\%$). Similarly, entangled states can also be stored by extending this procedure to either a pair of systems or bimodal cavities. The absence of electronic or spin-state interaction excludes a relevant source of losses, such as spontaneous atomic decay or dephasing. Our protocol works in a single-step operation based solely on transferring light states between two coupled resonators.

The results were achieved using input-output theory \cite{kuhn2012, raphaeli2012, baragiola2012}, which determines the output state by relating internal dynamics to a known input state from a bosonic reservoir \cite{walls}. Illustratively, we derived analytical solutions for a coupled set of Schrödinger and Heisenberg-Langevin equations, corresponding, respectively, to inputs of single-photon and coherent states. 
The key point to obtain those solutions is to impose no output states, which is in accordance with the phase-match condition. For this condition to be fulfilled, the time-dependent coupling of the system must be modulated in such a way that, after the input field fully enters the cavity mode $\text{B}$, the coupling is turned off, shutting down any interaction between this mode and the remaining system.

\textit{Model}---We investigate the dynamics of a system composed of two resonant coupled resonators, a two-sided cavity $\text{A}$ coupled to both bosonic reservoir $\text{C}$ and a single-sided cavity $\text{B}$, as presented in Fig.~\ref{model}(a). Considering the Hamiltonian operator to be in the \textit{white-noise} regime \cite{qnoise}, we write it in the interaction picture rotating at the resonant frequency of the cavities as ($\hbar =1$)
\begin{align}\label{H full}
    H(t)  = & \ i g(t)\left( ab^\dagger - a^\dagger b \right) + \int^{+\infty}_{-\infty} \mathrm{d}\omega \, \omega \, C^\dagger_{\omega}C_{\omega}  \nonumber \\ &+  \frac{i}{\sqrt{\pi}}\int^{+\infty}_{-\infty}\mathrm{d}\omega\sqrt{\kappa} \left(a^\dagger C_{\omega}  - a C^\dagger_{\omega}\right) ,
\end{align}
where $a$ ($a^\dagger$) and $b$ ($b^\dagger$) are the annihilation (creation) operators of the intracavity modes, and $C_{\omega}$ ($C_{\omega}^{\dagger}$) are the frequency-dependent annihilation (creation) operators of the cavity bosonic reservoir $\text{C}$ \cite{Loudon2000}. The cavity $\text{A}$ is coupled to its reservoir via the decay rate of the field amplitude $\kappa$, and to the cavity $\text{B}$ via the time-dependent coupling strength $g(t)$.

The key point of this work is to adjust the coupling time dependency in such a way that the incoming pulse is completely stored in mode $\text{B}$. To that end, we will solve the equations describing the system dynamics using \textit{input-output} theory, then the appropriate form of $g(t)$ can be found for both the quantum and quasi-classical fields under consideration.


\textit{Single-Photon State}---To exhibit storage of a quantum state, there is no better representative case than a single-photon state. In this scenario, if the dynamics is prescribed as a single excitation in the whole composite system, the Schrödinger equation can be solved exactly. Considering initially that the cavity modes are in vacuum states and the bosonic reservoir has a single excitation, the initial state of the whole system can be written as the simple tensor product state
\begin{equation}
    \ket{\Psi(0)} = \underbrace{\ket{0}_a\otimes\ket{0}_{b}}_{\mathrm{Cavities}}\ \otimes \underbrace{\ket{1}_{C}}_{\mathrm{Reservoir}},
\end{equation}
where $\ket{1}_{C} = \int^{+\infty}_{-\infty} \mathrm{d}\omega  \chi_{\text{in}}(\omega)C^{\dagger}_{\omega}\ket{0}_{C}$ describes the input field in a continuous-mode and square-normalized single-photon pulse. The Fourier transform of the spectral density function $\chi_{\text{in}}(\omega)$ defines the temporal shape of the impinging pulse $\alpha_{\text{in}}(t)$ \cite{Loudon2000}. Thus, for a single excitation, the general evolved state is (omitting the tensor product symbols, for simplicity)
\begin{align}
    \ket{\Psi(t)}  &=  c_a(t)\ket{1}_a\ket{0}_b\ket{0}_C 
    + c_b(t)\ket{0}_a\ket{1}_b\ket{0}_C  \nonumber\\ &+ \int_{-\infty}^{+\infty} \mathrm{d}\omega \chi(\omega,t)\ket{0}_a\ket{0}_bC^\dagger_{\omega}\ket{0}_C.
\end{align}

The Schr\"odinger equation $i\partial_{t}\ket{\Psi(t)}=H(t)\ket{\Psi(t)}$ for this evolved state yields a set of integro-differential equations that can be transformed (shown in Supplemental Material \cite{supp}) into a coupled set of two differential equations for $c_a(t)$ and $c_b(t)$
\begin{subequations}\label{SE_sol}
    \begin{align}
        \frac{\mathrm{d}{c}_a(t)}{\mathrm{d}t} &= g(t)c_b(t) - \kappa c_a(t) + \sqrt{2\kappa}\alpha_\text{in}(t),\label{pa dynamic}\\
        \frac{\mathrm{d}{c}_b(t)}{\mathrm{d}t} &= -g(t)c_a(t), \label{pb dynamic}
    \end{align}
\end{subequations}
bound by the input-output relation $\alpha_\text{out}(t) = \sqrt{2\kappa_a}c_a(t) - \alpha_\text{in}(t)$. These equations are not immediately solvable analytically unless we impose some condition to find the time-dependent coupling $g(t)$. Because we are interested in storing the input pulse in one of the modes, we impose $\alpha_\text{out}(t) = 0$, fulfilling both the mathematical and physical interests of our case supported by the phase-match condition. Under such a condition, it is clear from the input-output relation that $c_a(t) = \alpha_\text{in}(t)/\sqrt{2\kappa}$, hence the time-dependent coupling is 
\begin{equation}\label{gt single photon}
    g(t) = \frac{1}{\sqrt{2\kappa}}\left( \frac{\mathrm{d}}{\mathrm{d}t}\alpha_\text{in}(t)  - \kappa \alpha_\text{in}(t) \right)\frac{1}{c_b(t)}.
\end{equation}
with
\begin{equation}\label{a_sol_se}
    c_b(t) = \left(\int^t_{0}\alpha_\text{in}^{2}(t')\mathrm{d}t' - \frac{\alpha_\text{in}^{2}(t)}{2\kappa}\right)^{1/2}.
\end{equation}
Here, one can see that this coupling depends solely on the input pulse shape.
Since the input pulse $\alpha_\text{in} (t)$ is given, calculating the probability of occupation of the cavity $\text{B}$ over time is straightforward via $|c_b(t)|^2$. In addition, since $\alpha_\text{in}(t)$ is a square-normalized function with $\alpha_\text{in}(t \to \infty) = 0$, from Eq.~\eqref{a_sol_se} we have $|c_b(t\to\infty)|^2 = 1$ for any temporal shape, which means the complete storage of any single-photon pulse given the appropriate modulation of $g(t)$. 


\textit{Coherent States}---To store quasi-classical states in the form of a coherent propagating pulse, with any mean number of photons 
a case of great interest, e.g., in quantum batteries, we must follow the previous description and find, again, the optimal time-dependence for the cavity coupling. In this case, solving the Schrödinger equation as in the previous section is not convenient, unfortunately. That would imply a more complex time-evolved state, making the transition between integro-differential to purely differential equations a very laborious (if possible) task. However, one can obtain the time-dependent coupling using the Heisenberg-Langevin (HL) equations \cite{walls}, where the populations for each mode are obtained by calculating the dynamics of the mean values of the system operators instead of the probability amplitudes. 

We begin by defining the general input-output operators $c_{\text{in}}(t)$ and $c_{\text{out}}(t)$ as \cite{walls}
\begin{align}
    c_{\text{in}}(t) &= -\frac{1}{\sqrt{2 \pi}}\int^{+\infty}_{-\infty}\mathrm{d} \omega e^{-i \omega(t - t_{0})}C_\omega(t_0), \\
    c_{\text{out}}(t) &= \frac{1}{\sqrt{2 \pi}}\int^{+\infty}_{-\infty}\mathrm{d} \omega e^{-i \omega(t - t_{1})}C_\omega(t_1),
\end{align}
with $t_{0} \to -\infty$ and $t_{1}  \to +\infty$. The HL equations are written for any time-dependent operator $O(t)$ as \cite{walls}
\begin{align}\label{h-l}
    \frac{\mathrm{d}O(t)}{\mathrm{d}t} &= -i\left[O(t), H_{\text{sys}}(t)\right] \nonumber\\
    &- \left[O(t), a^{\dagger}(t)\right]\left(\kappa a(t) - \sqrt{2\kappa}c_{\text{in}}(t)\right) \nonumber\\
    &+\left(\kappa a^{\dagger}(t)-\sqrt{2\kappa}c_{\text{in}}(t)\right)\left[O(t),a(t)\right],
\end{align}
where $H_{\text{sys}}(t) = i g(t)\left( a(t)b^\dagger(t) - a^\dagger(t) b(t) \right)$. The equation above yields the general input-output relation $c_{\text{out}}(t) = \sqrt{2\kappa}a(t) - c_{\text{in}}(t)$, which is the boundary condition for the dynamical equations to be obtained.
For $a(t)$ and $b(t)$, we obtain the following set of differential equations
    \begin{subequations}
        \begin{align}\label{h-la}
            \frac{\mathrm{d}a(t)}{\mathrm{d}t} &= g(t) b(t) - \kappa a(t) + \sqrt{2\kappa}c_{\text{in}}(t), \\
            \frac{\mathrm{d}b(t)}{\mathrm{d}t} &= - g(t) a(t). \label{h-lb}
        \end{align}
\end{subequations}
Note that the equations above are very similar to the ones obtained in Eq.~\eqref{SE_sol}, yet they have a distinct physical meaning. 

We can solve Eqs.~(\ref{h-la}) and (\ref{h-lb}) as mean values evolving from the initial state, hence the solutions are obtained similarly to the single-photon case. Imposing this time $c_{\text{out}}(t)=0$, in accordance to the phase-match condition we find that the time-dependent coupling to be 
\begin{equation}\label{gt_hl}
g(t) = \frac{1}{\sqrt{2\kappa}}\left( \frac{\mathrm{d}}{\mathrm{d}t}\moy{c_\text{in}(t)} - \kappa \moy{c_\text{in}(t)}  \right)\frac{1}{\moy{b(t)}},
\end{equation}
%
with
 \begin{equation}\label{h-l_asol}
     \moy{b(t)} = \left( \int^t_{0}\moy{c^{2}_\text{in}(t')}\mathrm{d}t' -\frac{\moy{c^{2}_\text{in}(t)}}{2\kappa}  \right)^{1/2}.
 \end{equation}
In this way, we are finally able to access the population of the mode $\text{B}$ over time due to a coherent input field via the time-dependent number operator $\langle b^{\dagger}(t)b(t)\rangle = |\moy{b(t)}|^2$ given that $\langle c_\text{in}(t)\rangle=\sqrt{n_{p}}\alpha_\text{in}(t)$, where $n_{p}$ defines the mean number of photons distributed in the input pulse with temporal shape $\alpha_\text{in}(t)$. Similarly to the case of a single-photon pulse as the input field, from Eq.~\eqref{h-l_asol} we have $|\moy{b(t\to\infty)}|^2 = n_{p}$ for any temporal shape, which means complete storage of any coherent pulse given the appropriate modulation of $g(t)$.


\begin{figure}[t]
\begin{centering}
\includegraphics[trim = 0.7cm 0.35cm 1.2cm 1.1cm, width =0.49\columnwidth, clip]{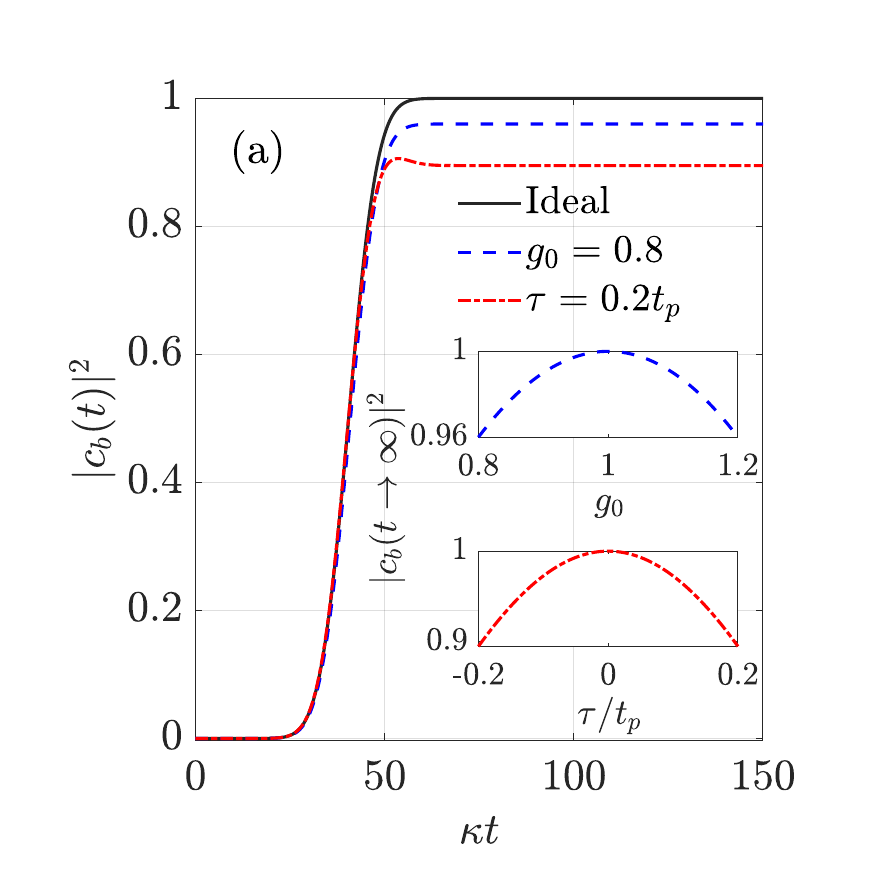}
\includegraphics[trim = 0.7cm 0.35cm 1.2cm 1.1cm, width =0.49\columnwidth,clip]{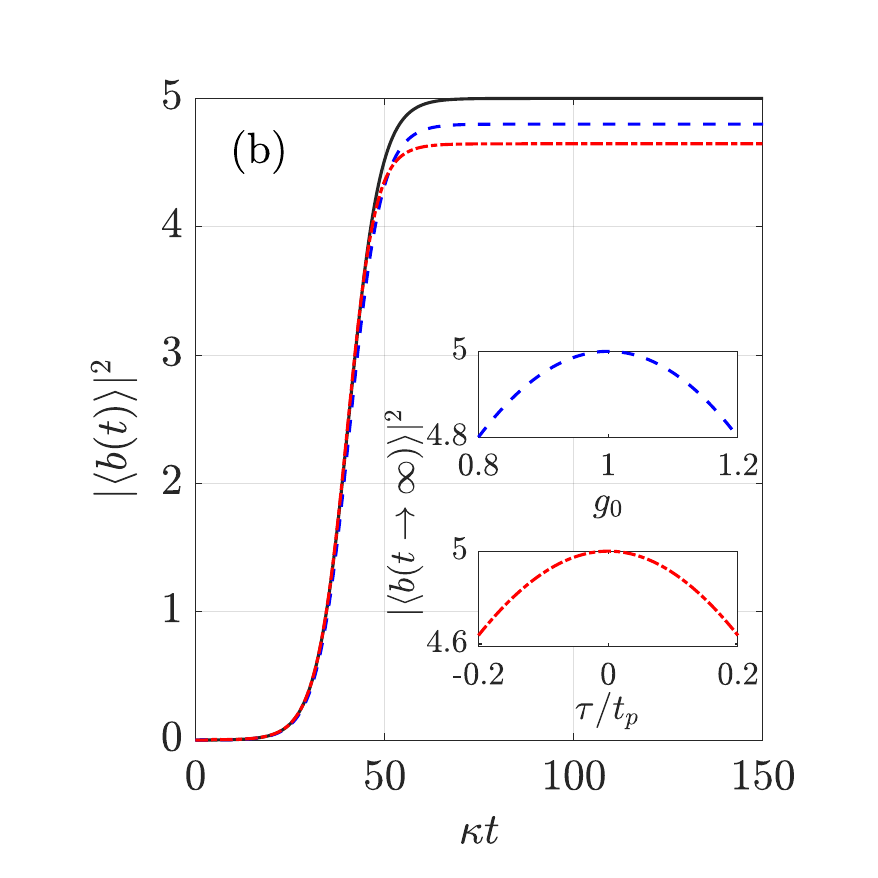}
\par\end{centering}
\caption{{Efficiency of the model.} (a) Single photon occupation probability of cavity $\text{B}$ over time considering the ideal case (solid line), coupling amplitude deviation (dashed line) and time delay (dash-dotted line). Here we set $\kappa t_p=20$ for a single-photon Gaussian pulse. (b) Mean number of photons over time for a coherent state with a mean number of photons $n_{p}=5$ distributed over a hyperbolic secant pulse. In both panels, the insets show, for the two proposed sources of error, how the probability scales at a sufficiently future time ($t \to \infty$) when there is no longer interaction between the two cavities.}
\label{pb plot}
\end{figure}

\textit{Discussions}---For an ideal case, the solid black line in Fig.~\ref{pb plot}(a) illustrates the complete storage of a single-photon state ($|c_b(t \to \infty)|^2=1$) considering the input field as a single-photon pulse with a Gaussian temporal shape $\alpha_{\text{in}}(t) = (\eta\sqrt{\pi})^{-1/2}\exp{\left[-{(t-T)^2}/{2\eta^2}\right]}$, where $T$ represents the time at which its maximum reaches cavity $\text{A}$ and $t_{p}=2\eta\sqrt{2\ln(2)}$ describes the pulse duration (full width at half maximum) with $\eta$ being the gaussian distribution standard deviation
. Assuming $t_{p}$ to be sufficiently long, we ensure that the spectral width fits within the cavity linewidth ($2\kappa$) \cite{borges2018}. Applying this $\alpha_{\text{in}}(t)$ to Eq.~\eqref{a_sol_se}, we find the following analytical solution
\begin{align}
    |c_{b}^{\text{(G)}}(t)|^2=\frac{\text{erf}\left(\frac{T}{\eta}\right)+\text{erf}\left(\frac{t-T}{\eta}\right)}{2}-\frac{e^{-\frac{(t-T)^2}{\eta^2}}}{2\sqrt{\pi}\kappa\eta},
\end{align}
plotted in Fig.~\ref{pb plot}(a). We emphasize that any square-normalized temporal shape can be considered for this system, for example, one can choose hyperbolic secant pulses, $\alpha_{\text{in}}(t) = \left(2\beta\right)^{-1/2}\text{sech}((t - T)/{\beta})$, which have been studied in superconducting systems for the last few years \cite{Ku2017,barron_2020}. For this case, the analytical solution can be obtained and written as 
\begin{align}
    |c^{(S)}_{b}(t)|^2 = \frac{\text{tanh}{\left(\frac{T}{\beta}\right)} + \text{tanh}{\left(\frac{t - T}{\beta}\right)}}{2} - \frac{\text{sech}^2(\frac{t - T}{\beta})}{4\kappa\beta}.
\end{align}
Analogous solutions can be obtained for Eq.~\eqref{h-l_asol}, related to the storage of coherent states, by applying $\alpha_{\text{in}}(t)\rightarrow \sqrt{n_{p}} \alpha_{\text{in}}(t)$, which leads to the same qualitative behavior shown in the solid line of Fig.~\ref{pb plot}(a), but with the population converging to the mean number of photons, $|\moy{b(t\to\infty)}|^2 = n_{p}$. As an example, the solid black line in Fig.~\ref{pb plot}(b) illustrates the complete storage of a coherent input field considering a hyperbolic secant pulse.

It is clear that the analytical solutions proposed here depend on an ideal setup. However, it is possible to analyze how different sources of errors impact our solutions by numerically solving Eqs.~\eqref{pa dynamic} and \eqref{pb dynamic}. Since our protocol relies mainly on modulating the coupling, these errors must be associated with it. In the next, two main sources are considered: the first is an arbitrary amplitude deviation in the coupling $g(t)\rightarrow g_{0} \ g^{(\text{I})}(t)$, where $g^{(\text{I})}(t)$ is the coupling in an ideal scenario. As shown by the dashed lines in Fig.~\ref{pb plot}, our model demonstrates robustness against this kind of error, evidencing a decrease in storage efficiency by as much as $4\%$ when $g_0$ changes by up to $20\%$. The second source of error is on the coupling time scale, because it is expected from a practical standpoint that modulating this coupling in terms of a propagating light pulse will be the hardest task. To that end, we can consider a delay (positive or negative) in the coupling time as $g(t)\rightarrow g(t-\tau)$. As expected, this significantly affects the efficiency of the system, where a delay of $20\%$ related to the pulse duration can decrease the storage efficiency to $90\%$, as seen in the dash-dotted line of Fig.~\ref{pb plot}. Most importantly, if these coupling imperfections are mitigated, take errors up to $5\%$ for example, the system still holds efficiency above $99.99\%$.


The procedure can be naturally extended to store two-mode \textit{entangled states}, such as those generated via the STIRAP technique \cite{weber2008}. This can be achieved by considering a pair of systems illustrated in Fig.~\ref{model}(a), where each mode interacts with one of the systems. Then, the state of each mode is stored in cavity B of its respective system. On the other hand, a two-mode entangled state can still be stored employing only a single system composed of cavities $\text{A}$ and $\text{B}$, provided that each cavity supports two orthogonally polarized modes \cite{wilk2007,rempe2015}. In this scenario, the entanglement must be between single-photon polarizations rather than between photon numbers.

\textit{Conclusions}----We proposed an optical system able to efficiently store both quantum (single-photon) and quasi-classical (coherent) states of light by accordingly adjusting the coupling between two resonant cavities. 
We have analytically shown solutions that are dependent only on the temporal shape of the input pulse, making the choice of it completely arbitrary if the phase-match condition is considered. Furthermore, we have shown numerically that even outside of an ideal setup, high efficiency is still held since the system is robust against coupling amplitude deviation and small time delays in the coupling control. Since efficient storage of light states is indispensable for both quantum memories and batteries, we believe that our scheme fits both purposes. In terms of coherent states, an envelope with any number of photons can be efficiently stored, making the system suitable for a battery charging model. On the other hand, single-photon or entangled states can be encoded on a qubit basis, such that this setup can be arranged as an efficient quantum memory, where the on-demand output from cavity $\text{B}$ can be achieved by reactivating the coupling between $\text{B}$ and $\text{A}$. Due to its simple structure, it is clear that our system can be implemented not only with optical cavities but also with superconducting systems \cite{matanin2023}. We acknowledge the hardship of generating such time-dependent coupling in any platform, but nonetheless, this scheme may pave new ways to manipulate quantum information.

\textit{Acknowledgments}---This work was supported by the Brazilian National Council for Scientific and Technological Development (CNPq, Grants Nos. 405712/2023-5, 311612/2021-0), CAPES-COFECUB (CAPES, Grants Nos. 88887.711967/2022-00), the Royal Society via a University Research Fellowship (URF/R1/201158) and by the S\~ao Paulo Research Foundation (FAPESP, Grants Nos. 2019/11999-5, 2018/22402-7, 2022/00209-6 and 2022/10218-2).


\bibliography{refs.bib}

\begin{thebibliography}{47}%
\makeatletter
\providecommand \@ifxundefined [1]{%
 \@ifx{#1\undefined}
}%
\providecommand \@ifnum [1]{%
 \ifnum #1\expandafter \@firstoftwo
 \else \expandafter \@secondoftwo
 \fi
}%
\providecommand \@ifx [1]{%
 \ifx #1\expandafter \@firstoftwo
 \else \expandafter \@secondoftwo
 \fi
}%
\providecommand \natexlab [1]{#1}%
\providecommand \enquote  [1]{``#1''}%
\providecommand \bibnamefont  [1]{#1}%
\providecommand \bibfnamefont [1]{#1}%
\providecommand \citenamefont [1]{#1}%
\providecommand \href@noop [0]{\@secondoftwo}%
\providecommand \href [0]{\begingroup \@sanitize@url \@href}%
\providecommand \@href[1]{\@@startlink{#1}\@@href}%
\providecommand \@@href[1]{\endgroup#1\@@endlink}%
\providecommand \@sanitize@url [0]{\catcode `\\12\catcode `\$12\catcode `\&12\catcode `\#12\catcode `\^12\catcode `\_12\catcode `\%12\relax}%
\providecommand \@@startlink[1]{}%
\providecommand \@@endlink[0]{}%
\providecommand \url  [0]{\begingroup\@sanitize@url \@url }%
\providecommand \@url [1]{\endgroup\@href {#1}{\urlprefix }}%
\providecommand \urlprefix  [0]{URL }%
\providecommand \Eprint [0]{\href }%
\providecommand \doibase [0]{https://doi.org/}%
\providecommand \selectlanguage [0]{\@gobble}%
\providecommand \bibinfo  [0]{\@secondoftwo}%
\providecommand \bibfield  [0]{\@secondoftwo}%
\providecommand \translation [1]{[#1]}%
\providecommand \BibitemOpen [0]{}%
\providecommand \bibitemStop [0]{}%
\providecommand \bibitemNoStop [0]{.\EOS\space}%
\providecommand \EOS [0]{\spacefactor3000\relax}%
\providecommand \BibitemShut  [1]{\csname bibitem#1\endcsname}%
\let\auto@bib@innerbib\@empty
\bibitem [{\citenamefont {Monroe}(2002)}]{monroe2002}%
  \BibitemOpen
  \bibfield  {author} {\bibinfo {author} {\bibfnamefont {C.}~\bibnamefont {Monroe}},\ }\bibfield  {title} {\bibinfo {title} {Quantum information processing with atoms and photons},\ }\href {https://doi.org/10.1038/416238a} {\bibfield  {journal} {\bibinfo  {journal} {Nature (London)}\ }\textbf {\bibinfo {volume} {416}},\ \bibinfo {pages} {238} (\bibinfo {year} {2002})}\BibitemShut {NoStop}%
\bibitem [{\citenamefont {Kimble}(2008)}]{kimbleqi}%
  \BibitemOpen
  \bibfield  {author} {\bibinfo {author} {\bibfnamefont {H.~J.}\ \bibnamefont {Kimble}},\ }\bibfield  {title} {\bibinfo {title} {The quantum internet},\ }\href {https://doi.org/10.1038/nature07127} {\bibfield  {journal} {\bibinfo  {journal} {Nature (London)}\ }\textbf {\bibinfo {volume} {453}},\ \bibinfo {pages} {1023} (\bibinfo {year} {2008})}\BibitemShut {NoStop}%
\bibitem [{\citenamefont {Simon}\ \emph {et~al.}(2010)\citenamefont {Simon}, \citenamefont {Afzelius}, \citenamefont {Appel}, \citenamefont {Boyer de~la Giroday}, \citenamefont {Dewhurst}, \citenamefont {Gisin}, \citenamefont {Hu}, \citenamefont {Jelezko}, \citenamefont {Kröll}, \citenamefont {Müller}, \citenamefont {Nunn}, \citenamefont {Polzik}, \citenamefont {Rarity}, \citenamefont {De~Riedmatten}, \citenamefont {Rosenfeld}, \citenamefont {Shields}, \citenamefont {Sköld}, \citenamefont {Stevenson}, \citenamefont {Thew}, \citenamefont {Walmsley}, \citenamefont {Weber}, \citenamefont {Weinfurter}, \citenamefont {Wrachtrup},\ and\ \citenamefont {Young}}]{Simon2010}%
  \BibitemOpen
  \bibfield  {author} {\bibinfo {author} {\bibfnamefont {C.}~\bibnamefont {Simon}}, \bibinfo {author} {\bibfnamefont {M.}~\bibnamefont {Afzelius}}, \bibinfo {author} {\bibfnamefont {J.}~\bibnamefont {Appel}}, \bibinfo {author} {\bibfnamefont {A.}~\bibnamefont {Boyer de~la Giroday}}, \bibinfo {author} {\bibfnamefont {S.~J.}\ \bibnamefont {Dewhurst}}, \bibinfo {author} {\bibfnamefont {N.}~\bibnamefont {Gisin}}, \bibinfo {author} {\bibfnamefont {C.~Y.}\ \bibnamefont {Hu}}, \bibinfo {author} {\bibfnamefont {F.}~\bibnamefont {Jelezko}}, \bibinfo {author} {\bibfnamefont {S.}~\bibnamefont {Kröll}}, \bibinfo {author} {\bibfnamefont {J.~H.}\ \bibnamefont {Müller}}, \bibinfo {author} {\bibfnamefont {J.}~\bibnamefont {Nunn}}, \bibinfo {author} {\bibfnamefont {E.~S.}\ \bibnamefont {Polzik}}, \bibinfo {author} {\bibfnamefont {J.~G.}\ \bibnamefont {Rarity}}, \bibinfo {author} {\bibfnamefont {H.}~\bibnamefont {De~Riedmatten}}, \bibinfo {author} {\bibfnamefont {W.}~\bibnamefont {Rosenfeld}}, \bibinfo {author}
  {\bibfnamefont {A.~J.}\ \bibnamefont {Shields}}, \bibinfo {author} {\bibfnamefont {N.}~\bibnamefont {Sköld}}, \bibinfo {author} {\bibfnamefont {R.~M.}\ \bibnamefont {Stevenson}}, \bibinfo {author} {\bibfnamefont {R.}~\bibnamefont {Thew}}, \bibinfo {author} {\bibfnamefont {I.~A.}\ \bibnamefont {Walmsley}}, \bibinfo {author} {\bibfnamefont {M.~C.}\ \bibnamefont {Weber}}, \bibinfo {author} {\bibfnamefont {H.}~\bibnamefont {Weinfurter}}, \bibinfo {author} {\bibfnamefont {J.}~\bibnamefont {Wrachtrup}},\ and\ \bibinfo {author} {\bibfnamefont {R.~J.}\ \bibnamefont {Young}},\ }\bibfield  {title} {\bibinfo {title} {Quantum memories: A review based on the {E}uropean integrated project “{Q}ubit {A}pplications ({QAP})”},\ }\href {https://doi.org/10.1140/epjd/e2010-00103-y} {\bibfield  {journal} {\bibinfo  {journal} {Eur. Phys. J. D}\ }\textbf {\bibinfo {volume} {58}},\ \bibinfo {pages} {1} (\bibinfo {year} {2010})}\BibitemShut {NoStop}%
\bibitem [{\citenamefont {Heshami}\ \emph {et~al.}(2016)\citenamefont {Heshami}, \citenamefont {England}, \citenamefont {Humphreys}, \citenamefont {Bustard}, \citenamefont {Acosta}, \citenamefont {Nunn},\ and\ \citenamefont {Sussman}}]{Heshami2016}%
  \BibitemOpen
  \bibfield  {author} {\bibinfo {author} {\bibfnamefont {K.}~\bibnamefont {Heshami}}, \bibinfo {author} {\bibfnamefont {D.~G.}\ \bibnamefont {England}}, \bibinfo {author} {\bibfnamefont {P.~C.}\ \bibnamefont {Humphreys}}, \bibinfo {author} {\bibfnamefont {P.~J.}\ \bibnamefont {Bustard}}, \bibinfo {author} {\bibfnamefont {V.~M.}\ \bibnamefont {Acosta}}, \bibinfo {author} {\bibfnamefont {J.}~\bibnamefont {Nunn}},\ and\ \bibinfo {author} {\bibfnamefont {B.~J.}\ \bibnamefont {Sussman}},\ }\bibfield  {title} {\bibinfo {title} {Quantum memories: emerging applications and recent advances},\ }\href {https://doi.org/10.1080/09500340.2016.1148212} {\bibfield  {journal} {\bibinfo  {journal} {J. Mod. Opt.}\ }\textbf {\bibinfo {volume} {63}},\ \bibinfo {pages} {2005} (\bibinfo {year} {2016})}\BibitemShut {NoStop}%
\bibitem [{\citenamefont {Lvovsky}\ \emph {et~al.}(2009)\citenamefont {Lvovsky}, \citenamefont {Sanders},\ and\ \citenamefont {Tittel}}]{lvovsky2009}%
  \BibitemOpen
  \bibfield  {author} {\bibinfo {author} {\bibfnamefont {A.}~\bibnamefont {Lvovsky}}, \bibinfo {author} {\bibfnamefont {B.}~\bibnamefont {Sanders}},\ and\ \bibinfo {author} {\bibfnamefont {W.}~\bibnamefont {Tittel}},\ }\bibfield  {title} {\bibinfo {title} {Optical quantum memory},\ }\href {https://doi.org/10.1038/nphoton.2009.231} {\bibfield  {journal} {\bibinfo  {journal} {Nat. Photonics}\ }\textbf {\bibinfo {volume} {3}},\ \bibinfo {pages} {706} (\bibinfo {year} {2009})}\BibitemShut {NoStop}%
\bibitem [{\citenamefont {Ma\^{\i}tre}\ \emph {et~al.}(1997)\citenamefont {Ma\^{\i}tre}, \citenamefont {Hagley}, \citenamefont {Nogues}, \citenamefont {Wunderlich}, \citenamefont {Goy}, \citenamefont {Brune}, \citenamefont {Raimond},\ and\ \citenamefont {Haroche}}]{maitre1997}%
  \BibitemOpen
  \bibfield  {author} {\bibinfo {author} {\bibfnamefont {X.}~\bibnamefont {Ma\^{\i}tre}}, \bibinfo {author} {\bibfnamefont {E.}~\bibnamefont {Hagley}}, \bibinfo {author} {\bibfnamefont {G.}~\bibnamefont {Nogues}}, \bibinfo {author} {\bibfnamefont {C.}~\bibnamefont {Wunderlich}}, \bibinfo {author} {\bibfnamefont {P.}~\bibnamefont {Goy}}, \bibinfo {author} {\bibfnamefont {M.}~\bibnamefont {Brune}}, \bibinfo {author} {\bibfnamefont {J.~M.}\ \bibnamefont {Raimond}},\ and\ \bibinfo {author} {\bibfnamefont {S.}~\bibnamefont {Haroche}},\ }\bibfield  {title} {\bibinfo {title} {Quantum memory with a single photon in a cavity},\ }\href {https://doi.org/10.1103/PhysRevLett.79.769} {\bibfield  {journal} {\bibinfo  {journal} {Phys. Rev. Lett.}\ }\textbf {\bibinfo {volume} {79}},\ \bibinfo {pages} {769} (\bibinfo {year} {1997})}\BibitemShut {NoStop}%
\bibitem [{\citenamefont {Giannelli}\ \emph {et~al.}(2018)\citenamefont {Giannelli}, \citenamefont {Schmit}, \citenamefont {Calarco}, \citenamefont {Koch}, \citenamefont {Ritter},\ and\ \citenamefont {Morigi}}]{Giannelli_2018}%
  \BibitemOpen
  \bibfield  {author} {\bibinfo {author} {\bibfnamefont {L.}~\bibnamefont {Giannelli}}, \bibinfo {author} {\bibfnamefont {T.}~\bibnamefont {Schmit}}, \bibinfo {author} {\bibfnamefont {T.}~\bibnamefont {Calarco}}, \bibinfo {author} {\bibfnamefont {C.~P.}\ \bibnamefont {Koch}}, \bibinfo {author} {\bibfnamefont {S.}~\bibnamefont {Ritter}},\ and\ \bibinfo {author} {\bibfnamefont {G.}~\bibnamefont {Morigi}},\ }\bibfield  {title} {\bibinfo {title} {Optimal storage of a single photon by a single intra-cavity atom},\ }\href {https://doi.org/10.1088/1367-2630/aae725} {\bibfield  {journal} {\bibinfo  {journal} {New Journal of Physics}\ }\textbf {\bibinfo {volume} {20}},\ \bibinfo {pages} {105009} (\bibinfo {year} {2018})}\BibitemShut {NoStop}%
\bibitem [{\citenamefont {Kollath-B\"onig}\ \emph {et~al.}(2024)\citenamefont {Kollath-B\"onig}, \citenamefont {Dellantonio}, \citenamefont {Giannelli}, \citenamefont {Schmit}, \citenamefont {Morigi},\ and\ \citenamefont {S\o{}rensen}}]{kollath_2024}%
  \BibitemOpen
  \bibfield  {author} {\bibinfo {author} {\bibfnamefont {J.~S.}\ \bibnamefont {Kollath-B\"onig}}, \bibinfo {author} {\bibfnamefont {L.}~\bibnamefont {Dellantonio}}, \bibinfo {author} {\bibfnamefont {L.}~\bibnamefont {Giannelli}}, \bibinfo {author} {\bibfnamefont {T.}~\bibnamefont {Schmit}}, \bibinfo {author} {\bibfnamefont {G.}~\bibnamefont {Morigi}},\ and\ \bibinfo {author} {\bibfnamefont {A.~S.}\ \bibnamefont {S\o{}rensen}},\ }\bibfield  {title} {\bibinfo {title} {Fast storage of photons in cavity-assisted quantum memories},\ }\href {https://doi.org/10.1103/PhysRevApplied.22.044038} {\bibfield  {journal} {\bibinfo  {journal} {Phys. Rev. Appl.}\ }\textbf {\bibinfo {volume} {22}},\ \bibinfo {pages} {044038} (\bibinfo {year} {2024})}\BibitemShut {NoStop}%
\bibitem [{\citenamefont {Arslanov}\ and\ \citenamefont {Moiseev}(2021)}]{arslanov2021}%
  \BibitemOpen
  \bibfield  {author} {\bibinfo {author} {\bibfnamefont {N.}~\bibnamefont {Arslanov}}\ and\ \bibinfo {author} {\bibfnamefont {S.}~\bibnamefont {Moiseev}},\ }\bibfield  {title} {\bibinfo {title} {Fast quantum memory on a single atom in a high-q cavity},\ }\href {https://doi.org/10.1007/s10946-021-09973-2} {\bibfield  {journal} {\bibinfo  {journal} {Journal of Russian Laser Research}\ }\textbf {\bibinfo {volume} {42}},\ \bibinfo {pages} {1} (\bibinfo {year} {2021})}\BibitemShut {NoStop}%
\bibitem [{\citenamefont {Kharlamova}\ \emph {et~al.}(2024)\citenamefont {Kharlamova}, \citenamefont {Arslanov},\ and\ \citenamefont {Moiseev}}]{kharlamova2024}%
  \BibitemOpen
  \bibfield  {author} {\bibinfo {author} {\bibfnamefont {Y.}~\bibnamefont {Kharlamova}}, \bibinfo {author} {\bibfnamefont {N.}~\bibnamefont {Arslanov}},\ and\ \bibinfo {author} {\bibfnamefont {S.}~\bibnamefont {Moiseev}},\ }\bibfield  {title} {\bibinfo {title} {Enhancing efficiency of the fast quantum memory on single-atom in cavity},\ }\href {https://doi.org/10.1134/S1063739723600747} {\bibfield  {journal} {\bibinfo  {journal} {Russian Microelectronics}\ }\textbf {\bibinfo {volume} {52}},\ \bibinfo {pages} {S395} (\bibinfo {year} {2024})}\BibitemShut {NoStop}%
\bibitem [{\citenamefont {Fleischhauer}\ and\ \citenamefont {Lukin}(2002)}]{fleischhauer2002}%
  \BibitemOpen
  \bibfield  {author} {\bibinfo {author} {\bibfnamefont {M.}~\bibnamefont {Fleischhauer}}\ and\ \bibinfo {author} {\bibfnamefont {M.~D.}\ \bibnamefont {Lukin}},\ }\bibfield  {title} {\bibinfo {title} {Quantum memory for photons: Dark-state polaritons},\ }\href {https://doi.org/10.1103/PhysRevA.65.022314} {\bibfield  {journal} {\bibinfo  {journal} {Phys. Rev. A}\ }\textbf {\bibinfo {volume} {65}},\ \bibinfo {pages} {022314} (\bibinfo {year} {2002})}\BibitemShut {NoStop}%
\bibitem [{\citenamefont {Moiseev}\ and\ \citenamefont {Tittel}(2011)}]{Moiseev2011}%
  \BibitemOpen
  \bibfield  {author} {\bibinfo {author} {\bibfnamefont {S.~A.}\ \bibnamefont {Moiseev}}\ and\ \bibinfo {author} {\bibfnamefont {W.}~\bibnamefont {Tittel}},\ }\bibfield  {title} {\bibinfo {title} {Optical quantum memory with generalized time-reversible atom–light interaction},\ }\href {https://doi.org/10.1088/1367-2630/13/6/063035} {\bibfield  {journal} {\bibinfo  {journal} {New J. Phys.}\ }\textbf {\bibinfo {volume} {13}},\ \bibinfo {pages} {063035} (\bibinfo {year} {2011})}\BibitemShut {NoStop}%
\bibitem [{\citenamefont {Novikova}\ \emph {et~al.}(2007)\citenamefont {Novikova}, \citenamefont {Gorshkov}, \citenamefont {Phillips}, \citenamefont {S\o{}rensen}, \citenamefont {Lukin},\ and\ \citenamefont {Walsworth}}]{novikova2007}%
  \BibitemOpen
  \bibfield  {author} {\bibinfo {author} {\bibfnamefont {I.}~\bibnamefont {Novikova}}, \bibinfo {author} {\bibfnamefont {A.~V.}\ \bibnamefont {Gorshkov}}, \bibinfo {author} {\bibfnamefont {D.~F.}\ \bibnamefont {Phillips}}, \bibinfo {author} {\bibfnamefont {A.~S.}\ \bibnamefont {S\o{}rensen}}, \bibinfo {author} {\bibfnamefont {M.~D.}\ \bibnamefont {Lukin}},\ and\ \bibinfo {author} {\bibfnamefont {R.~L.}\ \bibnamefont {Walsworth}},\ }\bibfield  {title} {\bibinfo {title} {Optimal control of light pulse storage and retrieval},\ }\href {https://doi.org/10.1103/PhysRevLett.98.243602} {\bibfield  {journal} {\bibinfo  {journal} {Phys. Rev. Lett.}\ }\textbf {\bibinfo {volume} {98}},\ \bibinfo {pages} {243602} (\bibinfo {year} {2007})}\BibitemShut {NoStop}%
\bibitem [{\citenamefont {Tittel}\ \emph {et~al.}(2010)\citenamefont {Tittel}, \citenamefont {Afzelius}, \citenamefont {Chaneliére}, \citenamefont {Cone}, \citenamefont {Kröll}, \citenamefont {Moiseev},\ and\ \citenamefont {Sellars}}]{tittel2010}%
  \BibitemOpen
  \bibfield  {author} {\bibinfo {author} {\bibfnamefont {W.}~\bibnamefont {Tittel}}, \bibinfo {author} {\bibfnamefont {M.}~\bibnamefont {Afzelius}}, \bibinfo {author} {\bibfnamefont {T.}~\bibnamefont {Chaneliére}}, \bibinfo {author} {\bibfnamefont {R.}~\bibnamefont {Cone}}, \bibinfo {author} {\bibfnamefont {S.}~\bibnamefont {Kröll}}, \bibinfo {author} {\bibfnamefont {S.}~\bibnamefont {Moiseev}},\ and\ \bibinfo {author} {\bibfnamefont {M.}~\bibnamefont {Sellars}},\ }\bibfield  {title} {\bibinfo {title} {Photon-echo quantum memory in solid state systems},\ }\href {https://doi.org/https://doi.org/10.1002/lpor.200810056} {\bibfield  {journal} {\bibinfo  {journal} {Laser Photonics Rev.}\ }\textbf {\bibinfo {volume} {4}},\ \bibinfo {pages} {244} (\bibinfo {year} {2010})}\BibitemShut {NoStop}%
\bibitem [{\citenamefont {Saglamyurek}\ \emph {et~al.}(2011)\citenamefont {Saglamyurek}, \citenamefont {Sinclair}, \citenamefont {Jin}, \citenamefont {Slater}, \citenamefont {Oblak}, \citenamefont {Bussières}, \citenamefont {George}, \citenamefont {Ricken}, \citenamefont {Sohler},\ and\ \citenamefont {Tittel}}]{Saglamyurek2011}%
  \BibitemOpen
  \bibfield  {author} {\bibinfo {author} {\bibfnamefont {E.}~\bibnamefont {Saglamyurek}}, \bibinfo {author} {\bibfnamefont {N.}~\bibnamefont {Sinclair}}, \bibinfo {author} {\bibfnamefont {J.}~\bibnamefont {Jin}}, \bibinfo {author} {\bibfnamefont {J.~A.}\ \bibnamefont {Slater}}, \bibinfo {author} {\bibfnamefont {D.}~\bibnamefont {Oblak}}, \bibinfo {author} {\bibfnamefont {F.}~\bibnamefont {Bussières}}, \bibinfo {author} {\bibfnamefont {M.}~\bibnamefont {George}}, \bibinfo {author} {\bibfnamefont {R.}~\bibnamefont {Ricken}}, \bibinfo {author} {\bibfnamefont {W.}~\bibnamefont {Sohler}},\ and\ \bibinfo {author} {\bibfnamefont {W.}~\bibnamefont {Tittel}},\ }\bibfield  {title} {\bibinfo {title} {Broadband waveguide quantum memory for entangled photons},\ }\href {https://doi.org/10.1038/nature09719} {\bibfield  {journal} {\bibinfo  {journal} {Nature (London)}\ }\textbf {\bibinfo {volume} {469}},\ \bibinfo {pages} {512} (\bibinfo {year} {2011})}\BibitemShut {NoStop}%
\bibitem [{\citenamefont {Lukin}(2003)}]{lukin2003}%
  \BibitemOpen
  \bibfield  {author} {\bibinfo {author} {\bibfnamefont {M.~D.}\ \bibnamefont {Lukin}},\ }\bibfield  {title} {\bibinfo {title} {Colloquium: Trapping and manipulating photon states in atomic ensembles},\ }\href {https://doi.org/10.1103/RevModPhys.75.457} {\bibfield  {journal} {\bibinfo  {journal} {Rev. Mod. Phys.}\ }\textbf {\bibinfo {volume} {75}},\ \bibinfo {pages} {457} (\bibinfo {year} {2003})}\BibitemShut {NoStop}%
\bibitem [{\citenamefont {Eisaman}\ \emph {et~al.}(2005)\citenamefont {Eisaman}, \citenamefont {André}, \citenamefont {Massou}, \citenamefont {Fleischhauer}, \citenamefont {Zibrov},\ and\ \citenamefont {Lukin}}]{eisaman2005}%
  \BibitemOpen
  \bibfield  {author} {\bibinfo {author} {\bibfnamefont {M.~D.}\ \bibnamefont {Eisaman}}, \bibinfo {author} {\bibfnamefont {A.}~\bibnamefont {André}}, \bibinfo {author} {\bibfnamefont {F.}~\bibnamefont {Massou}}, \bibinfo {author} {\bibfnamefont {M.}~\bibnamefont {Fleischhauer}}, \bibinfo {author} {\bibfnamefont {A.~S.}\ \bibnamefont {Zibrov}},\ and\ \bibinfo {author} {\bibfnamefont {M.~D.}\ \bibnamefont {Lukin}},\ }\bibfield  {title} {\bibinfo {title} {Electromagnetically induced transparency with tunable single-photon pulses},\ }\href {https://doi.org/10.1038/nature04327} {\bibfield  {journal} {\bibinfo  {journal} {Nature (London)}\ }\textbf {\bibinfo {volume} {438}},\ \bibinfo {pages} {837} (\bibinfo {year} {2005})}\BibitemShut {NoStop}%
\bibitem [{\citenamefont {Chanelière}\ \emph {et~al.}(2005)\citenamefont {Chanelière}, \citenamefont {Matsukevich}, \citenamefont {Jenkins}, \citenamefont {Lan}, \citenamefont {Kennedy},\ and\ \citenamefont {Kuzmich}}]{Chaneliere2005}%
  \BibitemOpen
  \bibfield  {author} {\bibinfo {author} {\bibfnamefont {T.}~\bibnamefont {Chanelière}}, \bibinfo {author} {\bibfnamefont {D.~N.}\ \bibnamefont {Matsukevich}}, \bibinfo {author} {\bibfnamefont {S.~D.}\ \bibnamefont {Jenkins}}, \bibinfo {author} {\bibfnamefont {S.-Y.}\ \bibnamefont {Lan}}, \bibinfo {author} {\bibfnamefont {T.~A.~B.}\ \bibnamefont {Kennedy}},\ and\ \bibinfo {author} {\bibfnamefont {A.}~\bibnamefont {Kuzmich}},\ }\bibfield  {title} {\bibinfo {title} {Storage and retrieval of single photons transmitted between remote quantum memories},\ }\href {https://doi.org/10.1038/nature04315} {\bibfield  {journal} {\bibinfo  {journal} {Nature (London)}\ }\textbf {\bibinfo {volume} {438}},\ \bibinfo {pages} {833} (\bibinfo {year} {2005})}\BibitemShut {NoStop}%
\bibitem [{\citenamefont {Chu}\ \emph {et~al.}(2024)\citenamefont {Chu}, \citenamefont {Lu}, \citenamefont {Chiang}, \citenamefont {Lin}, \citenamefont {Chen}, \citenamefont {Yu}, \citenamefont {Liao},\ and\ \citenamefont {Chen}}]{chu2024}%
  \BibitemOpen
  \bibfield  {author} {\bibinfo {author} {\bibfnamefont {K.-I.}\ \bibnamefont {Chu}}, \bibinfo {author} {\bibfnamefont {X.-C.}\ \bibnamefont {Lu}}, \bibinfo {author} {\bibfnamefont {K.-H.}\ \bibnamefont {Chiang}}, \bibinfo {author} {\bibfnamefont {Y.-H.}\ \bibnamefont {Lin}}, \bibinfo {author} {\bibfnamefont {C.-D.}\ \bibnamefont {Chen}}, \bibinfo {author} {\bibfnamefont {I.~A.}\ \bibnamefont {Yu}}, \bibinfo {author} {\bibfnamefont {W.-T.}\ \bibnamefont {Liao}},\ and\ \bibinfo {author} {\bibfnamefont {Y.-F.}\ \bibnamefont {Chen}},\ }\href@noop {} {\bibinfo {title} {{Slow and Stored Light via Electromagnetically Induced Transparency Using A $\Lambda$-type Superconducting Artificial Atom}}} (\bibinfo {year} {2024}),\ \Eprint {https://arxiv.org/abs/arXiv:2406.05007} {arXiv:2406.05007} \BibitemShut {NoStop}%
\bibitem [{\citenamefont {Pittman}\ and\ \citenamefont {Franson}(2002)}]{Pittman2002}%
  \BibitemOpen
  \bibfield  {author} {\bibinfo {author} {\bibfnamefont {T.~B.}\ \bibnamefont {Pittman}}\ and\ \bibinfo {author} {\bibfnamefont {J.~D.}\ \bibnamefont {Franson}},\ }\bibfield  {title} {\bibinfo {title} {Cyclical quantum memory for photonic qubits},\ }\href {https://doi.org/10.1103/PhysRevA.66.062302} {\bibfield  {journal} {\bibinfo  {journal} {Phys. Rev. A}\ }\textbf {\bibinfo {volume} {66}},\ \bibinfo {pages} {062302} (\bibinfo {year} {2002})}\BibitemShut {NoStop}%
\bibitem [{\citenamefont {Leung}\ and\ \citenamefont {Ralph}(2006)}]{leung2006}%
  \BibitemOpen
  \bibfield  {author} {\bibinfo {author} {\bibfnamefont {P.~M.}\ \bibnamefont {Leung}}\ and\ \bibinfo {author} {\bibfnamefont {T.~C.}\ \bibnamefont {Ralph}},\ }\bibfield  {title} {\bibinfo {title} {Quantum memory scheme based on optical fibers and cavities},\ }\href {https://doi.org/10.1103/PhysRevA.74.022311} {\bibfield  {journal} {\bibinfo  {journal} {Phys. Rev. A}\ }\textbf {\bibinfo {volume} {74}},\ \bibinfo {pages} {022311} (\bibinfo {year} {2006})}\BibitemShut {NoStop}%
\bibitem [{\citenamefont {Bouillard}\ \emph {et~al.}(2019)\citenamefont {Bouillard}, \citenamefont {Boucher}, \citenamefont {Ferrer~Ortas}, \citenamefont {Pointard},\ and\ \citenamefont {Tualle-Brouri}}]{bouillard2019}%
  \BibitemOpen
  \bibfield  {author} {\bibinfo {author} {\bibfnamefont {M.}~\bibnamefont {Bouillard}}, \bibinfo {author} {\bibfnamefont {G.}~\bibnamefont {Boucher}}, \bibinfo {author} {\bibfnamefont {J.}~\bibnamefont {Ferrer~Ortas}}, \bibinfo {author} {\bibfnamefont {B.}~\bibnamefont {Pointard}},\ and\ \bibinfo {author} {\bibfnamefont {R.}~\bibnamefont {Tualle-Brouri}},\ }\bibfield  {title} {\bibinfo {title} {Quantum storage of single-photon and two-photon fock states with an all-optical quantum memory},\ }\href {https://doi.org/10.1103/PhysRevLett.122.210501} {\bibfield  {journal} {\bibinfo  {journal} {Phys. Rev. Lett.}\ }\textbf {\bibinfo {volume} {122}},\ \bibinfo {pages} {210501} (\bibinfo {year} {2019})}\BibitemShut {NoStop}%
\bibitem [{\citenamefont {Evans}\ \emph {et~al.}(2023)\citenamefont {Evans}, \citenamefont {Nunn}, \citenamefont {Cheng}, \citenamefont {Franson},\ and\ \citenamefont {Pittman}}]{evans_2023}%
  \BibitemOpen
  \bibfield  {author} {\bibinfo {author} {\bibfnamefont {C.~J.}\ \bibnamefont {Evans}}, \bibinfo {author} {\bibfnamefont {C.~M.}\ \bibnamefont {Nunn}}, \bibinfo {author} {\bibfnamefont {S.~W.~L.}\ \bibnamefont {Cheng}}, \bibinfo {author} {\bibfnamefont {J.~D.}\ \bibnamefont {Franson}},\ and\ \bibinfo {author} {\bibfnamefont {T.~B.}\ \bibnamefont {Pittman}},\ }\bibfield  {title} {\bibinfo {title} {Experimental storage of photonic polarization entanglement in a broadband loop-based quantum memory},\ }\href {https://doi.org/10.1103/PhysRevA.108.L050601} {\bibfield  {journal} {\bibinfo  {journal} {Phys. Rev. A}\ }\textbf {\bibinfo {volume} {108}},\ \bibinfo {pages} {L050601} (\bibinfo {year} {2023})}\BibitemShut {NoStop}%
\bibitem [{\citenamefont {Alicki}\ and\ \citenamefont {Fannes}(2013)}]{alicki2012}%
  \BibitemOpen
  \bibfield  {author} {\bibinfo {author} {\bibfnamefont {R.}~\bibnamefont {Alicki}}\ and\ \bibinfo {author} {\bibfnamefont {M.}~\bibnamefont {Fannes}},\ }\bibfield  {title} {\bibinfo {title} {Entanglement boost for extractable work from ensembles of quantum batteries},\ }\href {https://doi.org/10.1103/PhysRevE.87.042123} {\bibfield  {journal} {\bibinfo  {journal} {Phys. Rev. E}\ }\textbf {\bibinfo {volume} {87}},\ \bibinfo {pages} {042123} (\bibinfo {year} {2013})}\BibitemShut {NoStop}%
\bibitem [{\citenamefont {Binder}\ \emph {et~al.}(2015)\citenamefont {Binder}, \citenamefont {Vinjanampathy}, \citenamefont {Modi},\ and\ \citenamefont {Goold}}]{Binder2015}%
  \BibitemOpen
  \bibfield  {author} {\bibinfo {author} {\bibfnamefont {F.~C.}\ \bibnamefont {Binder}}, \bibinfo {author} {\bibfnamefont {S.}~\bibnamefont {Vinjanampathy}}, \bibinfo {author} {\bibfnamefont {K.}~\bibnamefont {Modi}},\ and\ \bibinfo {author} {\bibfnamefont {J.}~\bibnamefont {Goold}},\ }\bibfield  {title} {\bibinfo {title} {Quantacell: powerful charging of quantum batteries},\ }\href {https://doi.org/10.1088/1367-2630/17/7/075015} {\bibfield  {journal} {\bibinfo  {journal} {New J. Phys.}\ }\textbf {\bibinfo {volume} {17}},\ \bibinfo {pages} {075015} (\bibinfo {year} {2015})}\BibitemShut {NoStop}%
\bibitem [{\citenamefont {Campaioli}\ \emph {et~al.}(2017)\citenamefont {Campaioli}, \citenamefont {Pollock}, \citenamefont {Binder}, \citenamefont {C\'eleri}, \citenamefont {Goold}, \citenamefont {Vinjanampathy},\ and\ \citenamefont {Modi}}]{campaioli2017}%
  \BibitemOpen
  \bibfield  {author} {\bibinfo {author} {\bibfnamefont {F.}~\bibnamefont {Campaioli}}, \bibinfo {author} {\bibfnamefont {F.~A.}\ \bibnamefont {Pollock}}, \bibinfo {author} {\bibfnamefont {F.~C.}\ \bibnamefont {Binder}}, \bibinfo {author} {\bibfnamefont {L.}~\bibnamefont {C\'eleri}}, \bibinfo {author} {\bibfnamefont {J.}~\bibnamefont {Goold}}, \bibinfo {author} {\bibfnamefont {S.}~\bibnamefont {Vinjanampathy}},\ and\ \bibinfo {author} {\bibfnamefont {K.}~\bibnamefont {Modi}},\ }\bibfield  {title} {\bibinfo {title} {Enhancing the charging power of quantum batteries},\ }\href {https://doi.org/10.1103/PhysRevLett.118.150601} {\bibfield  {journal} {\bibinfo  {journal} {Phys. Rev. Lett.}\ }\textbf {\bibinfo {volume} {118}},\ \bibinfo {pages} {150601} (\bibinfo {year} {2017})}\BibitemShut {NoStop}%
\bibitem [{\citenamefont {Campaioli}\ \emph {et~al.}(2024)\citenamefont {Campaioli}, \citenamefont {Gherardini}, \citenamefont {Quach}, \citenamefont {Polini},\ and\ \citenamefont {Andolina}}]{campioli2024_rev}%
  \BibitemOpen
  \bibfield  {author} {\bibinfo {author} {\bibfnamefont {F.}~\bibnamefont {Campaioli}}, \bibinfo {author} {\bibfnamefont {S.}~\bibnamefont {Gherardini}}, \bibinfo {author} {\bibfnamefont {J.~Q.}\ \bibnamefont {Quach}}, \bibinfo {author} {\bibfnamefont {M.}~\bibnamefont {Polini}},\ and\ \bibinfo {author} {\bibfnamefont {G.~M.}\ \bibnamefont {Andolina}},\ }\bibfield  {title} {\bibinfo {title} {Colloquium: Quantum batteries},\ }\href {https://doi.org/10.1103/RevModPhys.96.031001} {\bibfield  {journal} {\bibinfo  {journal} {Rev. Mod. Phys.}\ }\textbf {\bibinfo {volume} {96}},\ \bibinfo {pages} {031001} (\bibinfo {year} {2024})}\BibitemShut {NoStop}%
\bibitem [{\citenamefont {Zhang}\ \emph {et~al.}(2019)\citenamefont {Zhang}, \citenamefont {Yang}, \citenamefont {Fu},\ and\ \citenamefont {Wang}}]{zhang2019}%
  \BibitemOpen
  \bibfield  {author} {\bibinfo {author} {\bibfnamefont {Y.-Y.}\ \bibnamefont {Zhang}}, \bibinfo {author} {\bibfnamefont {T.-R.}\ \bibnamefont {Yang}}, \bibinfo {author} {\bibfnamefont {L.}~\bibnamefont {Fu}},\ and\ \bibinfo {author} {\bibfnamefont {X.}~\bibnamefont {Wang}},\ }\bibfield  {title} {\bibinfo {title} {Powerful harmonic charging in a quantum battery},\ }\href {https://doi.org/10.1103/PhysRevE.99.052106} {\bibfield  {journal} {\bibinfo  {journal} {Phys. Rev. E}\ }\textbf {\bibinfo {volume} {99}},\ \bibinfo {pages} {052106} (\bibinfo {year} {2019})}\BibitemShut {NoStop}%
\bibitem [{\citenamefont {Chen}\ \emph {et~al.}(2020)\citenamefont {Chen}, \citenamefont {Zhan}, \citenamefont {Shao}, \citenamefont {Zhang}, \citenamefont {Zhang},\ and\ \citenamefont {Wang}}]{Chen2020}%
  \BibitemOpen
  \bibfield  {author} {\bibinfo {author} {\bibfnamefont {J.}~\bibnamefont {Chen}}, \bibinfo {author} {\bibfnamefont {L.}~\bibnamefont {Zhan}}, \bibinfo {author} {\bibfnamefont {L.}~\bibnamefont {Shao}}, \bibinfo {author} {\bibfnamefont {X.}~\bibnamefont {Zhang}}, \bibinfo {author} {\bibfnamefont {Y.}~\bibnamefont {Zhang}},\ and\ \bibinfo {author} {\bibfnamefont {X.}~\bibnamefont {Wang}},\ }\bibfield  {title} {\bibinfo {title} {Charging quantum batteries with a general harmonic driving field},\ }\href {http://dx.doi.org/10.1002/andp.201900487} {\bibfield  {journal} {\bibinfo  {journal} {Ann. Phys.}\ }\textbf {\bibinfo {volume} {532}} (\bibinfo {year} {2020})}\BibitemShut {NoStop}%
\bibitem [{\citenamefont {Crescente}\ \emph {et~al.}(2020)\citenamefont {Crescente}, \citenamefont {Carrega}, \citenamefont {Sassetti},\ and\ \citenamefont {Ferraro}}]{Crescente2020}%
  \BibitemOpen
  \bibfield  {author} {\bibinfo {author} {\bibfnamefont {A.}~\bibnamefont {Crescente}}, \bibinfo {author} {\bibfnamefont {M.}~\bibnamefont {Carrega}}, \bibinfo {author} {\bibfnamefont {M.}~\bibnamefont {Sassetti}},\ and\ \bibinfo {author} {\bibfnamefont {D.}~\bibnamefont {Ferraro}},\ }\bibfield  {title} {\bibinfo {title} {Charging and energy fluctuations of a driven quantum battery},\ }\href {https://doi.org/10.1088/1367-2630/ab91fc} {\bibfield  {journal} {\bibinfo  {journal} {New J. Phys.}\ }\textbf {\bibinfo {volume} {22}},\ \bibinfo {pages} {063057} (\bibinfo {year} {2020})}\BibitemShut {NoStop}%
\bibitem [{\citenamefont {Downing}\ and\ \citenamefont {Ukhtary}(2024{\natexlab{a}})}]{Downing2024}%
  \BibitemOpen
  \bibfield  {author} {\bibinfo {author} {\bibfnamefont {C.~A.}\ \bibnamefont {Downing}}\ and\ \bibinfo {author} {\bibfnamefont {M.~S.}\ \bibnamefont {Ukhtary}},\ }\bibfield  {title} {\bibinfo {title} {Energetics of a pulsed quantum battery},\ }\href {https://doi.org/10.1209/0295-5075/ad2e79} {\bibfield  {journal} {\bibinfo  {journal} {Europhys. Lett.}\ }\textbf {\bibinfo {volume} {146}},\ \bibinfo {pages} {10001} (\bibinfo {year} {2024}{\natexlab{a}})}\BibitemShut {NoStop}%
\bibitem [{\citenamefont {Gemme}\ \emph {et~al.}(2024)\citenamefont {Gemme}, \citenamefont {Grossi}, \citenamefont {Vallecorsa}, \citenamefont {Sassetti},\ and\ \citenamefont {Ferraro}}]{gemme2024}%
  \BibitemOpen
  \bibfield  {author} {\bibinfo {author} {\bibfnamefont {G.}~\bibnamefont {Gemme}}, \bibinfo {author} {\bibfnamefont {M.}~\bibnamefont {Grossi}}, \bibinfo {author} {\bibfnamefont {S.}~\bibnamefont {Vallecorsa}}, \bibinfo {author} {\bibfnamefont {M.}~\bibnamefont {Sassetti}},\ and\ \bibinfo {author} {\bibfnamefont {D.}~\bibnamefont {Ferraro}},\ }\bibfield  {title} {\bibinfo {title} {Qutrit quantum battery: Comparing different charging protocols},\ }\href {https://doi.org/10.1103/PhysRevResearch.6.023091} {\bibfield  {journal} {\bibinfo  {journal} {Phys. Rev. Res.}\ }\textbf {\bibinfo {volume} {6}},\ \bibinfo {pages} {023091} (\bibinfo {year} {2024})}\BibitemShut {NoStop}%
\bibitem [{\citenamefont {Downing}\ and\ \citenamefont {Ukhtary}(2024{\natexlab{b}})}]{downing2024b}%
  \BibitemOpen
  \bibfield  {author} {\bibinfo {author} {\bibfnamefont {C.}~\bibnamefont {Downing}}\ and\ \bibinfo {author} {\bibfnamefont {M.}~\bibnamefont {Ukhtary}},\ }\bibfield  {title} {\bibinfo {title} {Two-photon charging of a quantum battery with a gaussian pulse envelope},\ }\href {https://doi.org/https://doi.org/10.1016/j.physleta.2024.129693} {\bibfield  {journal} {\bibinfo  {journal} {Phys. Lett. A}\ }\textbf {\bibinfo {volume} {518}},\ \bibinfo {pages} {129693} (\bibinfo {year} {2024}{\natexlab{b}})}\BibitemShut {NoStop}%
\bibitem [{\citenamefont {Dilley}\ \emph {et~al.}(2012)\citenamefont {Dilley}, \citenamefont {Nisbet-Jones}, \citenamefont {Shore},\ and\ \citenamefont {Kuhn}}]{kuhn2012}%
  \BibitemOpen
  \bibfield  {author} {\bibinfo {author} {\bibfnamefont {J.}~\bibnamefont {Dilley}}, \bibinfo {author} {\bibfnamefont {P.}~\bibnamefont {Nisbet-Jones}}, \bibinfo {author} {\bibfnamefont {B.~W.}\ \bibnamefont {Shore}},\ and\ \bibinfo {author} {\bibfnamefont {A.}~\bibnamefont {Kuhn}},\ }\bibfield  {title} {\bibinfo {title} {Single-photon absorption in coupled atom-cavity systems},\ }\href {https://doi.org/10.1103/PhysRevA.85.023834} {\bibfield  {journal} {\bibinfo  {journal} {Phys. Rev. A}\ }\textbf {\bibinfo {volume} {85}},\ \bibinfo {pages} {023834} (\bibinfo {year} {2012})}\BibitemShut {NoStop}%
\bibitem [{\citenamefont {Rephaeli}\ and\ \citenamefont {Fan}(2012)}]{raphaeli2012}%
  \BibitemOpen
  \bibfield  {author} {\bibinfo {author} {\bibfnamefont {E.}~\bibnamefont {Rephaeli}}\ and\ \bibinfo {author} {\bibfnamefont {S.}~\bibnamefont {Fan}},\ }\bibfield  {title} {\bibinfo {title} {Few-photon single-atom cavity qed with input-output formalism in fock space},\ }\href {https://doi.org/10.1109/JSTQE.2012.2196261} {\bibfield  {journal} {\bibinfo  {journal} {IEEE J. Sel. Top. Quant.}\ }\textbf {\bibinfo {volume} {18}},\ \bibinfo {pages} {1754} (\bibinfo {year} {2012})}\BibitemShut {NoStop}%
\bibitem [{\citenamefont {Baragiola}\ \emph {et~al.}(2012)\citenamefont {Baragiola}, \citenamefont {Cook}, \citenamefont {Bra\ifmmode~\acute{n}\else \'{n}\fi{}czyk},\ and\ \citenamefont {Combes}}]{baragiola2012}%
  \BibitemOpen
  \bibfield  {author} {\bibinfo {author} {\bibfnamefont {B.~Q.}\ \bibnamefont {Baragiola}}, \bibinfo {author} {\bibfnamefont {R.~L.}\ \bibnamefont {Cook}}, \bibinfo {author} {\bibfnamefont {A.~M.}\ \bibnamefont {Bra\ifmmode~\acute{n}\else \'{n}\fi{}czyk}},\ and\ \bibinfo {author} {\bibfnamefont {J.}~\bibnamefont {Combes}},\ }\bibfield  {title} {\bibinfo {title} {$n$-photon wave packets interacting with an arbitrary quantum system},\ }\href {https://doi.org/10.1103/PhysRevA.86.013811} {\bibfield  {journal} {\bibinfo  {journal} {Phys. Rev. A}\ }\textbf {\bibinfo {volume} {86}},\ \bibinfo {pages} {013811} (\bibinfo {year} {2012})}\BibitemShut {NoStop}%
\bibitem [{\citenamefont {Walls}\ and\ \citenamefont {Milburn}(2008)}]{walls}%
  \BibitemOpen
  \bibfield  {author} {\bibinfo {author} {\bibfnamefont {D.}~\bibnamefont {Walls}}\ and\ \bibinfo {author} {\bibfnamefont {G.}~\bibnamefont {Milburn}},\ }\href {https://books.google.com.br/books?id=LiWsc3Nlf0kC} {\emph {\bibinfo {title} {Quantum Optics}}}\ (\bibinfo  {publisher} {Springer Berlin Heidelberg},\ \bibinfo {year} {2008})\BibitemShut {NoStop}%
\bibitem [{\citenamefont {Gardiner}\ and\ \citenamefont {Zoller}(2010)}]{qnoise}%
  \BibitemOpen
  \bibfield  {author} {\bibinfo {author} {\bibfnamefont {C.}~\bibnamefont {Gardiner}}\ and\ \bibinfo {author} {\bibfnamefont {P.}~\bibnamefont {Zoller}},\ }\href {https://books.google.com.br/books?id=PygdvgAACAAJ} {\emph {\bibinfo {title} {Quantum Noise: A Handbook of Markovian and Non-Markovian Quantum Stochastic Methods with Applications to Quantum Optics}}}\ (\bibinfo  {publisher} {Springer-Verlag, Berlin},\ \bibinfo {year} {2010})\BibitemShut {NoStop}%
\bibitem [{\citenamefont {Loudon}(2000)}]{Loudon2000}%
  \BibitemOpen
  \bibfield  {author} {\bibinfo {author} {\bibfnamefont {R.}~\bibnamefont {Loudon}},\ }\href {https://books.google.com.br/books?id=AEkfajgqldoC} {\emph {\bibinfo {title} {The Quantum Theory of Light}}}\ (\bibinfo  {publisher} {Oxford University Press, Oxford},\ \bibinfo {year} {2000})\BibitemShut {NoStop}%
\bibitem [{sup()}]{supp}%
  \BibitemOpen
  \href@noop {} {\bibinfo {title} {See {Supplemental Material} at {[URL will be inserted by publisher]} for details.}}\BibitemShut {Stop}%
\bibitem [{\citenamefont {Borges}\ \emph {et~al.}(2018)\citenamefont {Borges}, \citenamefont {Rossatto}, \citenamefont {Luiz},\ and\ \citenamefont {Villas-Boas}}]{borges2018}%
  \BibitemOpen
  \bibfield  {author} {\bibinfo {author} {\bibfnamefont {H.~S.}\ \bibnamefont {Borges}}, \bibinfo {author} {\bibfnamefont {D.~Z.}\ \bibnamefont {Rossatto}}, \bibinfo {author} {\bibfnamefont {F.~S.}\ \bibnamefont {Luiz}},\ and\ \bibinfo {author} {\bibfnamefont {C.~J.}\ \bibnamefont {Villas-Boas}},\ }\bibfield  {title} {\bibinfo {title} {Heralded entangling quantum gate via cavity-assisted photon scattering},\ }\href {https://doi.org/10.1103/PhysRevA.97.013828} {\bibfield  {journal} {\bibinfo  {journal} {Phys. Rev. A}\ }\textbf {\bibinfo {volume} {97}},\ \bibinfo {pages} {013828} (\bibinfo {year} {2018})}\BibitemShut {NoStop}%
\bibitem [{\citenamefont {Ku}\ \emph {et~al.}(2017)\citenamefont {Ku}, \citenamefont {Long}, \citenamefont {Wu}, \citenamefont {Bal}, \citenamefont {Lake}, \citenamefont {Barnes}, \citenamefont {Economou},\ and\ \citenamefont {Pappas}}]{Ku2017}%
  \BibitemOpen
  \bibfield  {author} {\bibinfo {author} {\bibfnamefont {H.~S.}\ \bibnamefont {Ku}}, \bibinfo {author} {\bibfnamefont {J.~L.}\ \bibnamefont {Long}}, \bibinfo {author} {\bibfnamefont {X.}~\bibnamefont {Wu}}, \bibinfo {author} {\bibfnamefont {M.}~\bibnamefont {Bal}}, \bibinfo {author} {\bibfnamefont {R.~E.}\ \bibnamefont {Lake}}, \bibinfo {author} {\bibfnamefont {E.}~\bibnamefont {Barnes}}, \bibinfo {author} {\bibfnamefont {S.~E.}\ \bibnamefont {Economou}},\ and\ \bibinfo {author} {\bibfnamefont {D.~P.}\ \bibnamefont {Pappas}},\ }\bibfield  {title} {\bibinfo {title} {Single qubit operations using microwave hyperbolic secant pulses},\ }\href {https://doi.org/10.1103/PhysRevA.96.042339} {\bibfield  {journal} {\bibinfo  {journal} {Phys. Rev. A}\ }\textbf {\bibinfo {volume} {96}},\ \bibinfo {pages} {042339} (\bibinfo {year} {2017})}\BibitemShut {NoStop}%
\bibitem [{\citenamefont {Barron}\ \emph {et~al.}(2020)\citenamefont {Barron}, \citenamefont {Calderon-Vargas}, \citenamefont {Long}, \citenamefont {Pappas},\ and\ \citenamefont {Economou}}]{barron_2020}%
  \BibitemOpen
  \bibfield  {author} {\bibinfo {author} {\bibfnamefont {G.~S.}\ \bibnamefont {Barron}}, \bibinfo {author} {\bibfnamefont {F.~A.}\ \bibnamefont {Calderon-Vargas}}, \bibinfo {author} {\bibfnamefont {J.}~\bibnamefont {Long}}, \bibinfo {author} {\bibfnamefont {D.~P.}\ \bibnamefont {Pappas}},\ and\ \bibinfo {author} {\bibfnamefont {S.~E.}\ \bibnamefont {Economou}},\ }\bibfield  {title} {\bibinfo {title} {Microwave-based arbitrary cphase gates for transmon qubits},\ }\href {https://doi.org/10.1103/PhysRevB.101.054508} {\bibfield  {journal} {\bibinfo  {journal} {Phys. Rev. B}\ }\textbf {\bibinfo {volume} {101}},\ \bibinfo {pages} {054508} (\bibinfo {year} {2020})}\BibitemShut {NoStop}%
\bibitem [{\citenamefont {Weber}\ \emph {et~al.}(2009)\citenamefont {Weber}, \citenamefont {Specht}, \citenamefont {M\"uller}, \citenamefont {Bochmann}, \citenamefont {M\"ucke}, \citenamefont {Moehring},\ and\ \citenamefont {Rempe}}]{weber2008}%
  \BibitemOpen
  \bibfield  {author} {\bibinfo {author} {\bibfnamefont {B.}~\bibnamefont {Weber}}, \bibinfo {author} {\bibfnamefont {H.~P.}\ \bibnamefont {Specht}}, \bibinfo {author} {\bibfnamefont {T.}~\bibnamefont {M\"uller}}, \bibinfo {author} {\bibfnamefont {J.}~\bibnamefont {Bochmann}}, \bibinfo {author} {\bibfnamefont {M.}~\bibnamefont {M\"ucke}}, \bibinfo {author} {\bibfnamefont {D.~L.}\ \bibnamefont {Moehring}},\ and\ \bibinfo {author} {\bibfnamefont {G.}~\bibnamefont {Rempe}},\ }\bibfield  {title} {\bibinfo {title} {Photon-photon entanglement with a single trapped atom},\ }\href {https://doi.org/10.1103/PhysRevLett.102.030501} {\bibfield  {journal} {\bibinfo  {journal} {Phys. Rev. Lett.}\ }\textbf {\bibinfo {volume} {102}},\ \bibinfo {pages} {030501} (\bibinfo {year} {2009})}\BibitemShut {NoStop}%
\bibitem [{\citenamefont {Wilk}\ \emph {et~al.}(2007)\citenamefont {Wilk}, \citenamefont {Webster}, \citenamefont {Kuhn},\ and\ \citenamefont {Rempe}}]{wilk2007}%
  \BibitemOpen
  \bibfield  {author} {\bibinfo {author} {\bibfnamefont {T.}~\bibnamefont {Wilk}}, \bibinfo {author} {\bibfnamefont {S.~C.}\ \bibnamefont {Webster}}, \bibinfo {author} {\bibfnamefont {A.}~\bibnamefont {Kuhn}},\ and\ \bibinfo {author} {\bibfnamefont {G.}~\bibnamefont {Rempe}},\ }\bibfield  {title} {\bibinfo {title} {Single-atom single-photon quantum interface},\ }\href {https://doi.org/10.1126/science.1143835} {\bibfield  {journal} {\bibinfo  {journal} {Science}\ }\textbf {\bibinfo {volume} {317}},\ \bibinfo {pages} {488} (\bibinfo {year} {2007})}\BibitemShut {NoStop}%
\bibitem [{\citenamefont {Reiserer}\ and\ \citenamefont {Rempe}(2015)}]{rempe2015}%
  \BibitemOpen
  \bibfield  {author} {\bibinfo {author} {\bibfnamefont {A.}~\bibnamefont {Reiserer}}\ and\ \bibinfo {author} {\bibfnamefont {G.}~\bibnamefont {Rempe}},\ }\bibfield  {title} {\bibinfo {title} {Cavity-based quantum networks with single atoms and optical photons},\ }\href {https://doi.org/10.1103/RevModPhys.87.1379} {\bibfield  {journal} {\bibinfo  {journal} {Rev. Mod. Phys.}\ }\textbf {\bibinfo {volume} {87}},\ \bibinfo {pages} {1379} (\bibinfo {year} {2015})}\BibitemShut {NoStop}%
\bibitem [{\citenamefont {Matanin}\ \emph {et~al.}(2023)\citenamefont {Matanin}, \citenamefont {Gerasimov}, \citenamefont {Moiseev}, \citenamefont {Smirnov}, \citenamefont {Ivanov}, \citenamefont {Malevannaya}, \citenamefont {Polozov}, \citenamefont {Zikiy}, \citenamefont {Samoilov}, \citenamefont {Rodionov},\ and\ \citenamefont {Moiseev}}]{matanin2023}%
  \BibitemOpen
  \bibfield  {author} {\bibinfo {author} {\bibfnamefont {A.~R.}\ \bibnamefont {Matanin}}, \bibinfo {author} {\bibfnamefont {K.~I.}\ \bibnamefont {Gerasimov}}, \bibinfo {author} {\bibfnamefont {E.~S.}\ \bibnamefont {Moiseev}}, \bibinfo {author} {\bibfnamefont {N.~S.}\ \bibnamefont {Smirnov}}, \bibinfo {author} {\bibfnamefont {A.~I.}\ \bibnamefont {Ivanov}}, \bibinfo {author} {\bibfnamefont {E.~I.}\ \bibnamefont {Malevannaya}}, \bibinfo {author} {\bibfnamefont {V.~I.}\ \bibnamefont {Polozov}}, \bibinfo {author} {\bibfnamefont {E.~V.}\ \bibnamefont {Zikiy}}, \bibinfo {author} {\bibfnamefont {A.~A.}\ \bibnamefont {Samoilov}}, \bibinfo {author} {\bibfnamefont {I.~A.}\ \bibnamefont {Rodionov}},\ and\ \bibinfo {author} {\bibfnamefont {S.~A.}\ \bibnamefont {Moiseev}},\ }\bibfield  {title} {\bibinfo {title} {Toward highly efficient multimode superconducting quantum memory},\ }\href {https://doi.org/10.1103/PhysRevApplied.19.034011} {\bibfield  {journal} {\bibinfo  {journal} {Phys. Rev. Appl.}\ }\textbf {\bibinfo
  {volume} {19}},\ \bibinfo {pages} {034011} (\bibinfo {year} {2023})}\BibitemShut {NoStop}%
\end{thebibliography}%

\end{document}